\documentclass[%
 prl,
 superscriptaddress,
 amsmath,amssymb,
reprint,%
longbibliography,
]{revtex4-2}

\usepackage{hyperref}
\usepackage{amssymb}
\usepackage{aeguill}
\usepackage{graphicx}
\usepackage{dcolumn}
\usepackage{bm}

\usepackage{color}
\usepackage{url}
\usepackage{verbatim}
\usepackage{lineno}
\usepackage[utf8]{inputenc}
\usepackage{color}
\usepackage{lineno}
\usepackage{mathtools}
\usepackage{orcidlink}

\draft 

\begin{document}



\title{Effect of surface magnetism on the x-ray spectra of hollow atoms    } 



\author{P.~Dergham}
\affiliation {Institut des NanoSciences de Paris, CNRS, Sorbonne Université, F-75005 Paris, France}
\author{C.~Prigent}
\affiliation {Institut des NanoSciences de Paris, CNRS, Sorbonne Université, F-75005 Paris, France}
\author{Ch. V. Ahmad}
\affiliation {Institut des NanoSciences de Paris, CNRS, Sorbonne Université, F-75005 Paris, France} 
\author{A.~Brassart}
\affiliation {Institut des NanoSciences de Paris, CNRS, Sorbonne Université, F-75005 Paris, France}
\author{L.~Duval}
\affiliation {Institut des NanoSciences de Paris, CNRS, Sorbonne Université, F-75005 Paris, France}
\affiliation {Laboratoire Kastler Brossel, Sorbonne Universit\'e, CNRS, ENS-PSL Research University, Coll\`ege de France, F-75005 Paris, France}
\author{E.~Lamour\orcidlink{https://orcid.org/0000-0001-8467-2637}}
\affiliation {Institut des NanoSciences de Paris, CNRS, Sorbonne Université, F-75005 Paris, France}
\author{S.~Macé}
\affiliation {Institut des NanoSciences de Paris, CNRS, Sorbonne Université, F-75005 Paris, France}
\author{S.~Steydli}
\affiliation {Institut des NanoSciences de Paris, CNRS, Sorbonne Université, F-75005 Paris, France}
\author{E. M. Tosi}
\affiliation {Institut des NanoSciences de Paris, CNRS, Sorbonne Université, F-75005 Paris, France}
\author{D.~Vernhet}
\affiliation {Institut des NanoSciences de Paris, CNRS, Sorbonne Université, F-75005 Paris, France}
\author{M.~Trassinelli\orcidlink{https://orcid.org/0000-0003-4414-1801}}
\email[]{martino.trassinelli@insp.jussieu.fr}
\affiliation {Institut des NanoSciences de Paris, CNRS, Sorbonne Université, F-75005 Paris, France}

\date{\today}

\begin{abstract}
We present evidence that the detection of surface magnetism without the application of an external magnetic field is possible by studying the x-ray emission of highly charged ions interacting at grazing incidence with the sample. 
Measuring the 3 keV $n=2 \to 1$  transition in the interaction of Ar\textsuperscript{17+} with a nickel sample at various temperatures gives access to the $n=2$ ion population. The latter evolves from a half-filled to a full shell during the ferro-paramagnetic phase transition of the sample, with the ferromagnetic phase reflecting the filling blocking due to Pauli principle. 
This finding puts an end to a longstanding controversy arising from contradictory studies of ion--surface interaction using Auger spectroscopy.


\end{abstract}

\pacs{}

\maketitle 

Surface properties often exhibit unexpected characteristics compared to bulk properties, but their characterization may be quite difficult. 
The implementation of highly charged ions (HCI) as a probe of surface properties has been considered as an interesting method for decades \cite{Winter1999,Winter2002,Wilhelm2022}. When a slow (velocity < 1 a.u.) HCI approaches a surface, the Coulomb potential between the ion of charge $q$ and the surface drops below the work function ($W$) of the material. 
Consequently, a resonant transfer of surface electrons takes place toward highly excited states of the ion (with principal quantum numbers close to $n_c \sim q/\sqrt{2W}$ in atomic units, with $W$ of a few eV) until its neutralization \cite{Burgdorfer1991}. 
This occurs well above the surface (at a critical distance $d_c$ that can reach a few tens of angstroms).
Since electron transfer is much faster (fs) than de-excitation (ps), a hollow atom is formed with occupied high-$n$ shells while the inner ones remain empty \cite{Donets1983,Briand1990,Winter1999}.
As the hollow atom continues to approach until it eventually penetrates the material, additional more complex processes are involved.
The dynamics of interaction is simplified in grazing angle collisions. 
When the incident ion approaches the surface at an angle less than a characteristic critical angle \cite{Winter2001} (generally less than a few degrees for HCI), the ion is reflected by the surface and does not touch or penetrate the sample. 
This considerably reduces the contribution of all near-surface processes such as interatomic Coulombic decay \cite{Wilhelm2022} but also those occurring below the surface \cite{Briand1996a}.
Assuming these interaction conditions, the scenario can be roughly divided into two main steps. 
In the first, the hollow atom begins to relax through mainly autoionization processes, which in parallel allow electrons to be recaptured from the surface (when the approach distance is $< d_c$) into ion states with $n \lesssim n_c$ and so on until the inner shells are filled up to neutralization. This occurs on a time scale of $\sim 1-10$ ps \cite{dEtat1993,Wilhelm2022,Werl2025}. 
In a second step, at the very end of the cascade, Auger emission competes with radiative decay to fill the innermost shells ($n = 3$, 2 and 1). 
As shown by Winecki et al \cite{Winecki1996} for argon ions, when the M-shell is half occupied, filling of the L and K shells takes place on a time scale of a few tens of fs. 
The signature of this last step can be found in the emitted x rays whose energy depends on the number of spectator electrons present at the time of emission. For example, the authors of \cite{dEtat1993,Briand1996} have demonstrated, using x-ray spectroscopy, that the L-shell filling of the HCI projectile approaching the sample is strongly dependent on the metallic or insulating nature of the surface.  
These studies naturally raise the question: Can x-ray spectroscopy of HCI-surface interactions be extended to the study of surface magnetism?

In grazing-angle collisions, since the dominant involved processes depicted above (capture of first-layer electrons, radiative de-excitation and Auger process) preserve, even partially, the electron spin alignment, the post-interaction ion can be expected to carry the signature of the magnetic order of the sample surface, if any. Early experiments on the magnetism of iron surfaces were mainly dedicated to polarization studies, revealing spin-polarized electrons in Auger KLL emission in \cite{Pfandzelter2001} or polarized visible-light emission in \cite{Winter1989,Narmann1991,Narmann1998,Unipan2005}.
In particular, in the studies of Närmann et al \cite{Narmann1998}, a different transition temperature of the iron surface magnetism was even observed in comparison to the bulk one.
However, all of these measurements require the sample to be polarized by the application of an external magnetic field.

Later, Unipan and collaborators developed a new method that did not require external magnetic fields.
It was based on measuring the energy of the Auger electron emission from slow HCI colliding unpolarized nickel (and iron) surfaces \cite{Unipan2006a,Unipan2006}.
Due to the typically small size of the sample portion involved in electron capture for each single ion (in the order of a few nm for incidence angles around $20^\circ$) compared to the size of the magnetic domains (a few $\mu$m or more in nickel), the ion captures electrons from a sample portion with a well-defined magnetic orientation, selectively populating (in terms of total spin) the excited level of the ion.
With incident He\textsuperscript{2+} ions, the observed temperature dependence of the relative intensities of the two Auger KLL components (a singlet-dominant peak versus a triplet-dominant one) was interpreted as a signature of sensitivity to surface magnetism switching from the ferromagnetic to the paramagnetic phase.
Several years later, Busch \textit{et al.} \cite{Busch2008, Busch2011} revisited this approach. The authors pointed out that oxygen contamination could explain the change of the work function ($W)$ with temperature and, consequently, the evolution of the intensities of the Auger transitions \footnote{More precisely, Busch and collaborators claimed that the intensity changes observed by Unipan \textit{et al.} were not caused by changes of the sample magnetic order but most probably by the effects stemming from desorption and dissolution processes of surface oxygen contamination and the consequent change of the work function of the sample.}, ruling out the possibility of probing surface magnetism with this spectroscopy technique.
Nevertheless, the enigma remains, as it is well established that nickel exhibits magnetic properties at the surface (e.g., from inverse photoemission \cite{Donath1990}, circular dichroism measurements \cite{VanderLaan1992} and theoretical predictions\cite{Mittendorfer1999}). 

Twelve years after the Unipan--Busch controversy, in this paper we present clear experimental evidence for the possibility of detecting the magnetism of the uppermost surface layer using highly charged ions as a probe.  
Our methodology closely follows the strategy adopted in previous works, grazing collision conditions and absence of an external magnetic field, but with an important change in the probe signal: the use of x-ray spectroscopy of the characteristic emission of ions during the interaction instead of Auger emission.
Unlike electron emission, the recorded x-rays are unambiguously emitted from the probing ion (and not from the sample), and correspond to a signal from the end of the cascade following multi-electron capture, which initially takes place in highly excited $n_c$ states. Consequently, a small variation in the preferential $n_c$ capture level due to a variation in $W$ will not influence the x-ray energy spectrum.
More specifically, we investigated the interaction between a 170 keV Ar\textsuperscript{17+} ion beam (with a K-shell vacancy giving rise to x-ray radiation around 3 keV) and a nickel sample ($W=5.01$ eV) at various temperatures.


The Ar\textsuperscript{17+} ion beam of a few nA was produced by the SIMPA facility (French acronym for Source d'Ions Multichargés de PAris) \cite{Gumberidze2010} which consists of an Electron-Cyclotron Resonance Ion Source (ECRIS) connected to an ultra-high vacuum beam-line to bring the ions into a dedicated interaction chamber with a typical pressure of 10\textsuperscript{-9} mbar (with the ion beam in the chamber). This chamber was equipped with a six-axis sample manipulator with high precision control over sample position, orientation, and temperature.
To visualize the ion beam and control its shape, two other silicon and ceramic targets were placed below the main sample location.
The fluorescence produced on these targets was detected by a CCD camera positioned at an angle of 30$^\circ$ to the beam axis. 
To measure the x rays emitted during ion--surface interactions, a silicon-drift detector \footnote{Rayspec x-ray detector model 881-3711-1A} was installed at an angle of 60$^\circ$.
To limit the region seen by the detector around the sample, and thus reduce background noise, a collimator was set up.
The detector was energy-calibrated using the emission from the silicon target and the silicon escape peak of the detector itself.

The sample consisted of a fcc nickel monocristal produced by the Surface Preparation Laboratory (the Nederlands) \cite{SPL} with a 10 mm diameter and a 2 mm thickness, and with the (110) plane on top.
As this type of experiment is still sensitive to contamination, the surface was systematically cleaned and checked before each ion-sample interaction. For this purpose, a preparation chamber with a typical pressure of 10\textsuperscript{-10} mbar was connected to the interaction chamber. There, the sample was cleaned by sputtering with an Ar\textsuperscript{+} beam for 30 minutes at an energy of 1 keV. 
In addition, reconstruction of the surface was obtained by flash annealing to $\sim 600$$^\circ$C \footnote{This operation was done in a regular base after the collision with the ions to limit the contamination form the heating preparation chamber elements before the interaction with the ions.}. 
Following the cleaning process, the absence of contaminants was checked by Auger electron spectroscopy of the sample surface just before transfer to the interaction chamber. 



To prevent penetration of the ions into the sample, the incidence angle of the ion beam was set to be smaller than the characteristic critical angle \cite{Winter2002}, equal to $3.23^\circ$ in our case. The relative angle between the goniometer encoder and the ion-beam direction was obtained by the x-ray yields resulting from the interaction of the ion beam with the silicon target as a function of the silicon surface angle (see Ref.~\cite{Dergham2022} for more details). 
Subsequently, a laser reflection was used to determine the angle offset between the orientation of the silicon target and the nickel sample. 
With this method, the ion--sample angle was adjusted to $0.9 \pm 0.3^\circ$. 

Following the preparation of the sample and the ion beam, series of x-ray spectra were acquired as a function of the sample temperature, \textit{i.e.} changing the magnetic phase of the sample from a ferromagnetic to a paramagnetic phase characterized by a Curie temperature of $T_\mathrm{C}= 354^\circ$C. 
However, in our setup, the maximum achievable temperature was limited to 252$^\circ$C due to degradation of the x-ray detector resolution at higher temperatures. As the detector window is transparent to infrared radiation, its spectral resolution at 3 keV increases from 130 eV at room temperature to more than 180 eV at 252$^\circ$C \footnote{Note that the resolution was checked at each temperature thanks to the presence of low-energy fluorescence lines.}).


\begin{figure}
    \centering
    \includegraphics[width=\columnwidth]{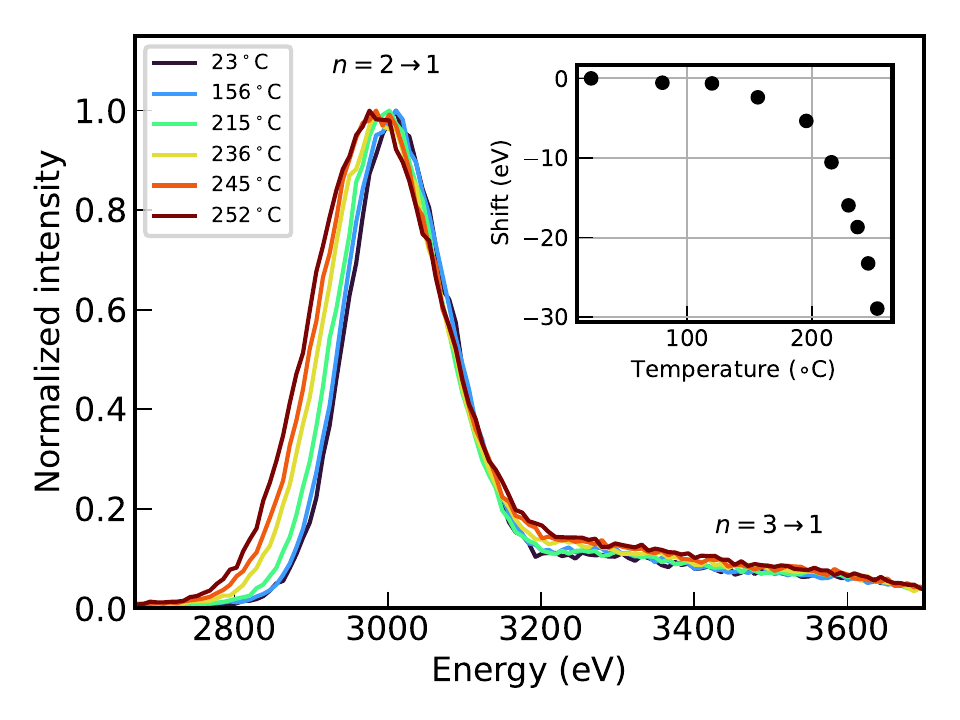}
    \caption{X-ray spectra for Ar\textsuperscript{17+} at 170 keV hitting the nickel sample for various temperatures. In the inset is reported the main line barycentre shift with respect to the room temperature spectrum, calculated considering the [2700 eV, 3200 eV] interval.}
    \label{fig:spectrum}
\end{figure}

\begin{figure}
    \centering
    \includegraphics[width=\columnwidth]{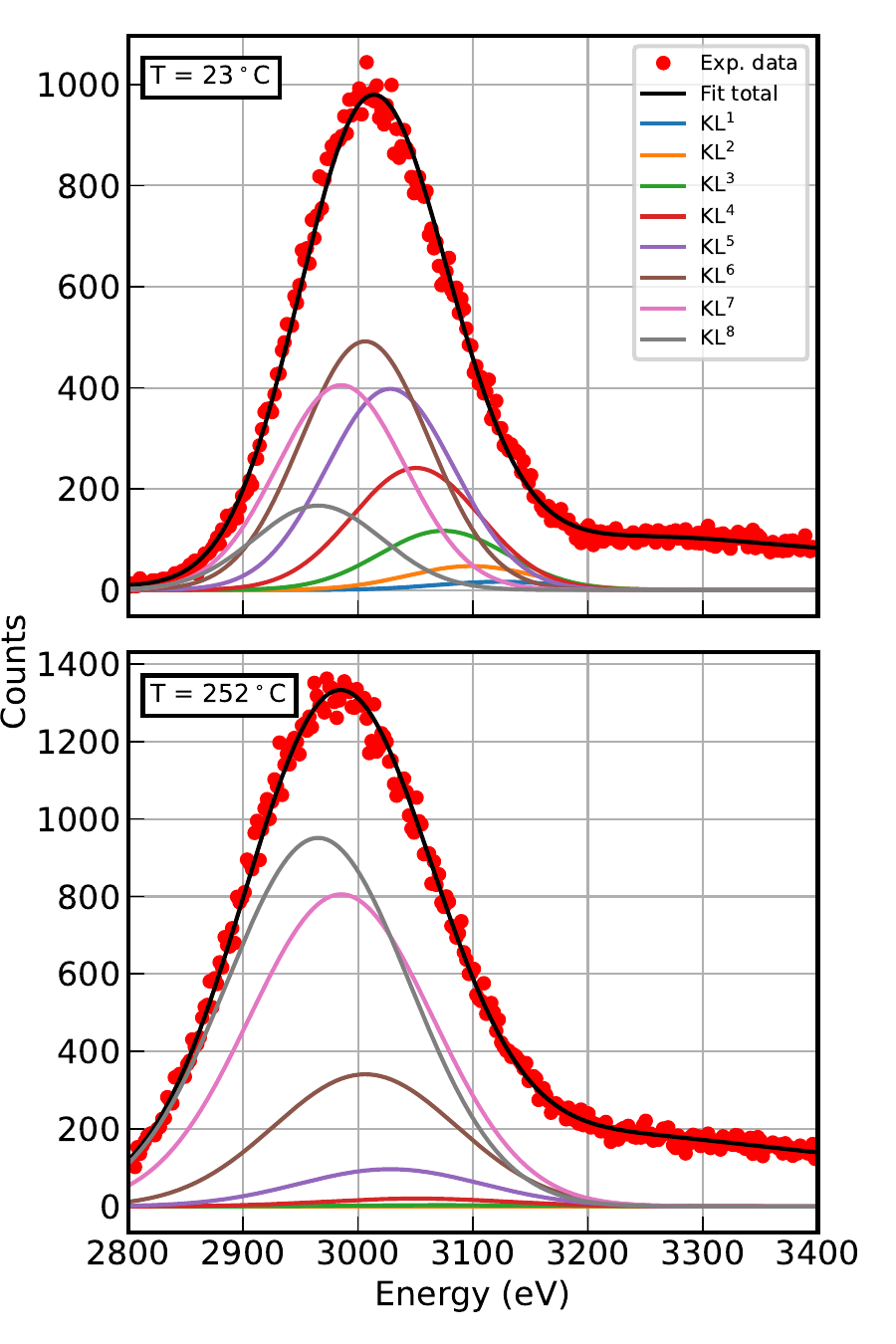}
    \caption{Spectral decomposition of the $n=2 \to 1$ x-ray transitions with 8 components (with $0 \le v_\mathrm{L} \le 7$ corresponding to  $8 \ge x \ge 1$) for the room temperature (top)  and $T = 252$$^\circ$C (bottom) measurements. Here, an occupancy $y = 0$ for the $n = 3$ level is assumed. For sake of clarity, the eight component intensities are enhanced by 70\%.
    }
    \label{fig:decomposition}
\end{figure}

When an Ar$^{17+}$ approaches the nickel sample surface, multi-electron capture in the $n_c \approx 18$ states occurs.
As already mentioned, the subsequent decay of such highly excited states leads, at the end of the cascade, to the x-ray emission recorded by our detector.
X-ray spectra obtained at different temperatures are presented in Fig.~\ref{fig:spectrum}.
The main 3 keV line corresponds to the unresolved group of the $1s\ 2\ell^x n'\ell'^y\to 1s^2\ 2\ell^{x-1}n'\ell'^y$ transitions, in short $n=2 \to 1$, also indicated using the spectroscopic KL$^x$ notation, where $1 \le x\le8$ relates to the number of electrons in the $n=2$ shell (see Fig.~\ref{fig:decomposition}). 
Note that at higher energy, the contribution of $n = 3 \to 1$ transitions is also visible in Fig.~\ref{fig:spectrum}. At first glance, it is noticeable that the main spectral component undergoes a shift towards low energies accompanied by a broadening as the temperature increases.
The barycenter shift of the $n=2 \to 1$ line is reported in the inset of Fig.~\ref{fig:spectrum}. As can be seen, the barycenter peak exhibits a discernible shift of almost 30 eV for sample temperatures between 23 and 252$^\circ$C. This shift can be simply interpreted by the gradual presence of additional electrons in the $n=2$ shell when the temperature increases. It has to be compared to the expected value taking into account the change of magnetic phases.
Two extreme cases can be considered. In the ideal case of fully polarized electron capture from a ferromagnetic domain (low temperature) and assuming no spin-misalignment during the radiative cascade decay of the ion, a maximum filling of $x=4$ electrons is expected due to Pauli exclusion principle.

To be able to extract the intensity of the single x-ray contributions for different electron fillings in the $n=2$ shell, further analysis is required considering both the energy shift and the evolution of the detector resolution.
Thus, for a given temperature, a decomposition of the main spectral peak into eight Gaussian profiles of the same width was carried out to model the elementary $1s\ 2\ell^x 3\ell'^y\to 1s^2\ 2\ell^{x-1}3\ell'^y$ components.
The energies of the eight components have been deduced from an empirical formula, similar to the one implemented by Winecki et al. \cite{Winecki1996} but also based on the values from Bhalla's tables \cite{Bhalla1973} with some refinement to reproduce emissions from He-like to $K_\alpha$ decay lines.
More precisely, we implemented the following formula valid for $x=1$ to 8 and $y=0$ to 8.
\begin{multline}
    E_{K \alpha} = 3154.7 -26.61 \, x  -6.44 \, y \\
    + 0.467 \, x \, y + 0.438 \, x^2 + 0.132 \, y^2 \text{ eV}.
\end{multline}

The energy proximity of the different components (about 30 eV, much smaller than the detector resolution) requires to consider an intensity distribution to simulate the spectra. 
A natural choice is to assume a Poisson distribution for the number of vacancies $v_\mathrm{L} = 8 - x$ in the L-shell
\begin{equation}
    I_X(v_\mathrm{L}) \propto \mu_\mathrm{L}^{v_\mathrm{L}} e^{- \mu_\mathrm{L}} / v_\mathrm{L}!
\end{equation}
where $\mu_\mathrm{L}$ is the average of the distribution. 
It is worth noting that a Poisson distribution describes qualitatively well the hollow atom spectra recorded with a higher resolving power (see, e.g. \cite{Briand1996,Briand2009}).


\begin{figure}
    \centering
    \includegraphics[width=\columnwidth]{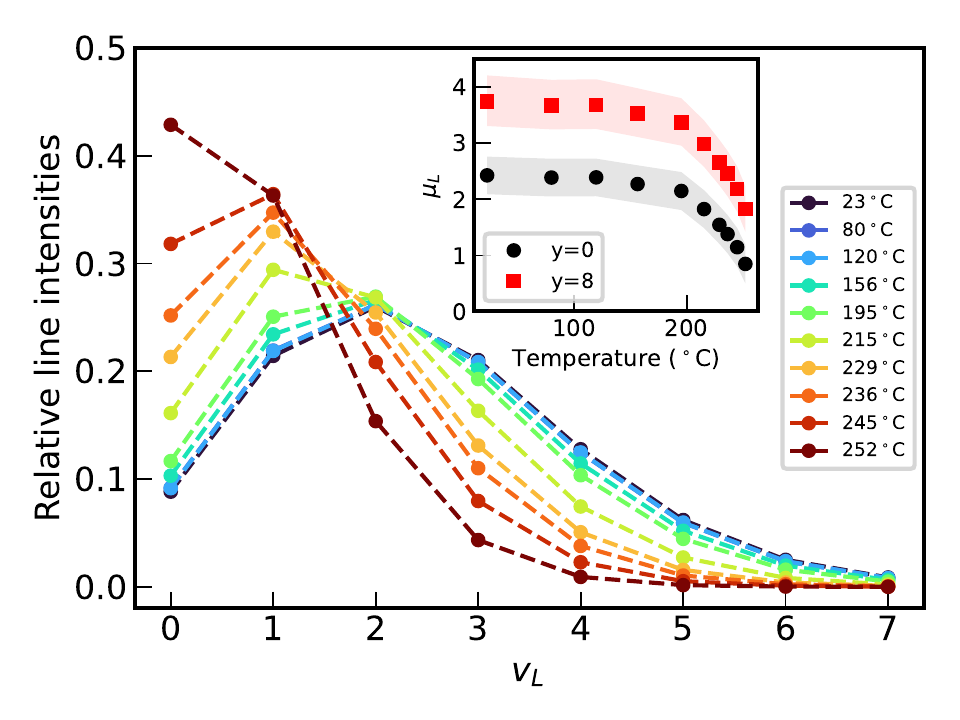}
    \caption{Relative intensities of the 8 components used to fit the $K_\alpha$ lines for each temperature assuming a Poisson distribution for their evolution and $y=0$ electrons in the $n=3$ levels.
    In the inset, evolutions with the sample temperature of the extracted mean values $\mu_\mathrm{L}$ of the number of apparent vacancies $v_\mathrm{L}$ in the $n=2$ shell assuming either $y=0$ or 8 electrons in the $n=3$ shell. The filled areas correspond to the uncertainty.}
    \label{fig:intensities}
\end{figure}

The influence of the population in the M-shell has been quantified using two extreme filling condition. 
The results of our analysis are presented in Figs.~\ref{fig:decomposition} and \ref{fig:intensities}.
In Fig.~\ref{fig:decomposition}, the spectral decomposition with $y=0$ is presented in detail for two temperatures (room temperature and $T = 252^\circ$C) taking into account the Doppler effect for each of the eight components, as well as the influence of the line $n=3 \to 1$ in the high-energy tail of the line $n=2 \to 1$. Clearly, the weight of each of the KL$^x$ intensities depends drastically on the sample temperature. 
Fig.~\ref{fig:intensities} summarizes all the relative intensities of KL$^x$ lines obtained from the spectral decompositions for each sample temperature assuming a Poisson distribution and $y=0$. From there, the average value $\mu_\mathrm{L}$ of \textit{apparent} vacancies $v_\mathrm{L}$ in the $n=2$ level can be extracted. 
While the KL$^6$M$^0$ peak (corresponding to $v_\mathrm{L}=2$) is the most intense at $T = 23^\circ$ C, the KL$^8$M$^0$ ($v_\mathrm{L}=0$) dominates at $T = 252^\circ$C.
The evolution of the $v_\mathrm{L}$ value with temperature is given in the inset of Fig.~\ref{fig:intensities} for two cases of occupancy $y$ (0 or 8) in the $n=3$ level. The associated uncertainty is mainly due to the detector calibration accuracy of about 7~eV for energies around 3000~eV. When a larger number of electrons $y$ in $n=3$ is considered, higher $\mu_\mathrm{L}$ values are found for compensating the screening effect of these electrons that induces a lower K$\alpha$ transition energies. However, regardless of the assumed value of $y$, the trend with temperature is similar. To complete our analysis, when the fluorescence yield is taken into account (from \cite{Larkins1971}), the mean value of the \emph{corrected} vacancies $\mu_\mathrm{Lc}$ in the L shell is smaller than $\mu_\mathrm{L}$ as reported in Table 1 for two temperatures and the two extreme occupancies of the M shell ($y=0$ and 8). 

\begin{table}
\caption{\label{tab:mu}
Average values of the observed $\mu_\mathrm{L}$ and corrected $\mu_{Lc}$ vacancies for the two extreme temperatures and different filling $y$ of the M shell.
}
\begin{ruledtabular}
\begin{tabular}{c|cc|cc}
 & \multicolumn{2}{c|}{Obs. vacancies $\mu_\mathrm{L}$} &\multicolumn{2}{c}{Corr. vacancies $\mu_\mathrm{Lc}$}\\
\colrule
 Temperature  & $y=0$ & $y=8$ & $y=0$ & $y=8$ \\
\colrule
$23^\circ$C & $2.4 \pm 0.3$ & $3.7 \pm 0.5$ & $2.0 \pm 0.3$ & $2.2 \pm 0.5$\\
$252^\circ$C & $0.8 \pm 0.3$ & $1.8 \pm 0.4$ & $0.7 \pm 0.3$ & $1.7 \pm 0.4$ \\
\end{tabular}
\end{ruledtabular}
\end{table}

Beyond these considerations, the lowest values of $\mu_\mathrm{Lc}$ (or $\mu_\mathrm{L}$) at high temperature correspond to almost complete filling of the L shell.
This observation agrees with previous high-resolution x-ray spectra of hollow atoms produced in proximity of clean paramagnetic metallic surfaces \cite{Briand2009} for which a small value of vacancies is induced by the mobility of valence electrons.
At room temperature, when the sample is ferromagnetic, larger values are found.
This shows that, beside the mobility of the electrons of the sample, their spin orientation plays a role in preventing the $n = 2$ shell from being completely filled by Pauli principle.

What may be surprising in our observations is that the $n = 2$ shell is almost entirely occupied at a temperature of 253~$^\circ$C, well below the Curie temperature $T_\mathrm{C}$=~350$^\circ$C. It is as if the ferro-paramagnetic phase transition at the surface occurs before reaching $T_\mathrm{C}$, the reference transition temperature for bulk. 
Nevertheless, a very different phase transition temperature value between surface and bulk has already been observed in the past in the study of Fe(110) surface magnetism \cite{Narmann1998}. There, the phase transition temperature at the surface was found at almost half of the absolute Curie temperature for bulk.
Our observations suggest that a similar behavior could occur for nickel, as also suggested in recent theoretical studies of nanowires magnetism of iron and nickel \cite{Courtes2024}. 


In conclusion, we present here evidence for the detection of surface magnetism from the x-ray emission of highly charged ions interacting  at grazing incidence with a sample.
Our results clarify a longstanding debate within the community due to conflicting measurements based on Auger spectroscopy. 
The difficulty associated with Auger spectroscopy, and in particular its dependence on the work function, is here circumvented by detecting photons emitted by the ion rather than electrons emitted by the ion or the sample.
The energy shift of the $n=2 \to 1$ peak observed when Ar\textsuperscript{17+} ions interact with a nickel sample at various temperatures is simply interpreted by a different filling of the $n=2$ shell.
At room temperature, electron capture from ferromagnetic domains results in only partial filling of this level due to the Pauli exclusion principle. 
At higher temperature, towards the paramagnetic phase of the sample, the loss of coherence in the spin alignment of captured electrons allows higher values of filling, corresponding to a shift of the $n=2 \to 1$ transition to lower energies. 
An in-depth  analysis based on a common intensity distribution of the eight KL$^x$ transitions that compose the main $n=2 \to 1$ spectral peak allows us to precisely track the evolution of the population in $n=2$ with the sample temperature.
Thus, at high temperature, a maximum filling is observed, in agreement with high-resolution x-ray spectra recorded using metallic surfaces \cite{Briand2009}.
As electron capture by the ion occurs well above the surface, the magnetic signature observed via x-ray emission comes from the very first atomic layer of the sample. 
Such a selective sensitivity opens up perspectives for the detection of topological magnetic states, as in 2D magnetic materials, without the need to apply external magnetic fields.
In the future, further details on the population of the atomic levels could be obtained by implementing a high-resolution x-ray spectrometer to resolve the KL$^x$ components. 
This will enable the $n = 2$ population to be quantified more precisely by determining accurate intensity ratios of the individual components. In addition, fine energy analysis of these lines will give access to the number of spectator electrons in the $n = 3$ shell at the moment of X-ray emission
This information will consequently give access to the timing of the different level filling and decay, as discussed in Ref.~\cite{dEtat1993}.


\begin{acknowledgments}
We thank J.-P. Rozet for all the knowledge he has shared with us. We would like to thank F. Aumayr, R. Whilhelm and M. Werl for the enriching discussions. 
This work has been partially supported by the French Agence Nationale de la Recherche (ANR) under reference ANR-20-CE91-0007 and the Austrian Fonds zur Förderung der wissenschaftlichen Forschung (FWF) under project No. I 4914-N (project DIMAS).
The data that support the findings of this article are not publicly available upon publication because it is not technically feasible and/or the cost of preparing, depositing, and hosting the data would be prohibitive within the terms of this research project. The data are available from the authors upon reasonable request. 

\end{acknowledgments}

\bibliography{ion-surf}

\end{document}